\documentclass[aps,prd,twocolumn,showpacs,superscriptaddress,groupedaddress,nofootinbib,floatfix]{revtex4-1}

\usepackage{amsmath}
\usepackage{amsfonts}
\usepackage{amssymb}	
\usepackage{graphicx}
\usepackage{bm}
\usepackage{color}
\usepackage{commath}
\allowdisplaybreaks

\def\GMc2{G M_{\odot} c^{-2}}

\def\lm{{\ell m}}

\def\lm{{\ell m}}

\def\lm{{\ell m}}

\newcommand\be{\begin{equation}}
\newcommand\ee{\end{equation}}

\def\TEOBResumS{\texttt{TEOBResumS}}

\usepackage[normalem]{ulem} 

\begin{document}

\title{Strong-field scattering of two spinning black holes: Numerics versus Analytics}

\author{Seth \surname{Hopper}${}^{1}$}
\author{Alessandro \surname{Nagar}${}^{2,3}$}
\author{Piero \surname{Rettegno}${}^{2,4}$}
\affiliation{${}^{1}$ Department of Physics \& Astronomy, Earlham College, 801 National Road West, Richmond, IN 47374, USA}
\affiliation{${}^2$ INFN Sezione di Torino, Via P. Giuria 1, 10125 Torino, Italy}
\affiliation{${}^3$ Institut des Hautes Etudes Scientifiques, 91440 Bures-sur-Yvette, France}
\affiliation{${}^{4}$ Dipartimento di Fisica, Universit\`a di Torino, via P. Giuria 1, 10125 Torino, Italy}

\begin{abstract}
We present new Numerical Relativity calculations of the scattering angle $\chi$ between
two, equal-mass, black holes on  hyperbolic-like orbits. We build upon previous work considering,
for the first time, spinning black holes, with equal spins either aligned or antialigned with the orbital angular momentum. We detail the numerical techniques used in the computation of $\chi$. 
Special care is taken in estimating error uncertainties on the quantities computed. The numerical values 
are compared with analytical predictions obtained using a new, state-of-the-art, effective one body model valid on 
generic orbits that incorporates post-Newtonian analytic information up to 5PN in the nonspinning, conservative 
sector and that has been additionally informed by Numerical Relativity simulations of quasi-circular coalescing black 
hole binaries. Our results indicate that the spin sector of the analytic model should be improved further in order 
to achieve satisfactory consistency with the most relativistic spinning configurations.
\end{abstract}  

\date{\today}

\maketitle

\section{Introduction}

Motivated by recent observational hints for eccentric and hyperbolic binary black hole (BBH) mergers in some LIGO-Virgo 
events \cite{LIGOScientific:2020iuh,Gayathri:2020coq,Gamba:2021gap} and by future prospect of detecting binary sources on generic orbits with 
LISA~\cite{LISA:2017pwj,Babak:2017tow,Gair:2017ynp}, the analytical relativity community is intensifying 
the efforts in building up accurate gravitational-wave (GW) models for {\it noncircularized} coalescing compact binaries~\cite{Hinder:2017sxy,Hinderer:2017jcs,Chiaramello:2020ehz,Loutrel:2020kmm,Nagar:2020xsk,Islam:2021mha,
Nagar:2021gss,Nagar:2021xnh,Albanesi:2021rby,Liu:2021pkr,Yun:2021jnh,Tucker:2021mvo,
Setyawati:2021gom,Khalil:2021txt,Ramos-Buades:2021adz,Placidi:2021rkh}.
The effective-one-body (EOB) framework~\cite{Buonanno:1998gg,Buonanno:2000ef,Damour:2000we,Damour:2001tu,Damour:2015isa} 
offers a theoretically complete method for a unified description of different classes of astrophysical binaries, from the 
comparable masses to the intermediate and extreme mass-ratio regimes, and for incorporating the 
fast motion and ringdown dynamics. 

Among the noncircular configurations, the case of hyperbolic scattering and hyperbolic capture 
is the one computationally more challenging. On the analytical side, after the pioneering work of 
Ref.~\cite{East:2012xq}, Ref.~\cite{Nagar:2020xsk} gave a comprehensive treatment of the phenomenon 
within the EOB model, constructing a complete waveform, through merger and ringdown, that was used 
in the analysis of GW190521~\cite{Gamba:2021gap}. 
The above-mentioned model was then improved in Refs.~\cite{Nagar:2021gss,Nagar:2021xnh}.
At a more fundamental level, the calculation of the gravitational scattering angle has received more and more
attention in recent years in the context of the formulation of an EOB model making use of the post-Minkowskian 
approximation~\cite{Damour:2016gwp,Damour:2017zjx,Damour:2019lcq,Bern:2019nnu,Damour:2020tta,Kalin:2020fhe,Bini:2020rzn,Bini:2021gat,Bern:2021yeh,Dlapa:2021vgp,Dlapa:2021npj,Manohar:2022dea,Khalil:2022ylj,Buonanno:2022pgc}.

On the Numerical Relativity (NR) side, after the pioneering work of Gold and Bruegmann~\cite{Gold:2012tk} on elliptic inspirals and 
hyperbolic encounters, Ref.~\cite{Damour:2014afa} presented the first, and so far only, computation of the 
scattering angle $\chi$ from equal-mass nonspinning black holes (BHs) on hyperbolic-like orbits. Scattering configurations 
were also considered in Ref.~\cite{Nelson:2019czq}, although the focus of the paper was on the spin-up of the two 
BHs induced by the close encounter and no calculation of $\chi$ was reported.
In this paper we extend the parameter space explored in Ref.~\cite{Damour:2014afa}, presenting the computation
of the scattering angle for 32 new equal-mass, equal-spin (with spin aligned with the orbital angular momentum) 
configurations. Of these 32 dataset, only 5 are nonspinning. 
In Ref.~\cite{Damour:2014afa} the initial energy of the configurations was constant and the starting angular momentum varied.
Here, for each value of the spin, we (approximately) fix the 
initial angular momentum and progressively increase the energy.
Similarly to Ref.~\cite{Damour:2014afa}, the numerical scattering angles are compared with the analytical calculation 
obtained using the state-of-the-art {\tt TEOBResumS} model for generic configurations of Ref.~\cite{Nagar:2021xnh}.

This paper is organized as follows. In Sec.~\ref{sec:NR} we describe our NR simulations, focusing
in particular on the computation of the scattering angle from the extrapolation of the puncture trajectories and
on the calculation of the energy and angular momentum losses during the process. The calculation of the scattering angle is 
reported in Sec.~\ref{sec:eobnr}, where it is also compared with the EOB predictions.
Our conclusions are collected in Sec.~\ref{conclusions}. 

We use geometrized units with $c=G=1$.

\section{Numerical Relativity: scattering angle, waveforms and fluxes}
\label{sec:NR}
We report here the results of a large number of NR simulation performed at the
beginning of 2016 and never published since. They were done using the 
November 2015 (Sommerville) release of the the Einstein Toolkit~\cite{Loffler:2011ay,EinsteinToolkit}. 
As such, the techniques used here are similar to those behind the results of Ref.~\cite{Damour:2014afa},
and thus the current paper should be seen as a complement to the information reported there.
\begin{table*}[t]
	\caption{\label{tab:ID} Initial data for the NR simulations. From left to right the columns report: configuration number; dimensionless spin; $(p_x,p_y)$ 
	in-plane components of the  momentum $\vec{p}_1$ of the $m_1$ BH located at $x=X=50M$ (note that $\vec{p_2}=-\vec{p}_1$); initial ADM
	energy, initial ADM total angular momentum and initial ADM orbital angular momentum.}
	\begin{center}
		\begin{ruledtabular}
			\begin{tabular}{c c c c c l c  c} 
				$\#$  & $S_i/m_i^2$ & $p_x/M$ & $p_y/M$ & $E_{\rm ADM}/M$ & $J_{\rm ADM}/M^2 $ & $J_{\rm ADM}^{\rm orb}/M^2$ \\	
				\hline
				\hline												
				1         & 0        &   $-0.05942506$	 & 0.01151993      &  1.00457189170793	& 1.15198647944995& 1.15198647944995\\									
				2         & 0        &   $-0.09423272$	 & 0.01151978  &   1.0148091006975      & 1.15195577447483 &1.15195577447483 \\											
				3         & 0        &   $-0.10755541$	& 0.01151963   &    1.01992105354885     & 1.15192562845173  &1.15192562845173\\	
				4          &0        &   $-0.11943931$	& 0.01151938   &     1.02502400428497    & 1.15187555184956  &1.15187555184956\\											
				5          &0        &    $-0.14030562$	& 0.01151787 &      1.03513920185028   &  1.15157506725085 &1.15157506725085\\											
				\hline
		                s1         & 0.6   &   $-0.11576611$ &   0.01216040    &  1.07665374989219       & 1.6179219954769    &  1.28364059145275\\	
		                s2          &0.6    &   $-0.12796939$  &  0.01215992  & 1.08202144718546        & 1.61779453645181  & 1.28353946693697 \\
		                s3          &0.6    &  $-0.13914499$  &   0.01215957  &  1.08740041845793      & 1.6177018945377     & 1.2834659659142\\
		                s4          &0.6    &  $-0.14952098$  &    0.01215909 &  1.09276714425576    &  1.61757391942674 &1.28336443194187 \\
		                s5          &0.6    &   $-0.16845199$ &    0.01215769 &  1.10345914737548   & 1.61720025417128  &1.28306797025159 \\
		                s6          &0.6    &   $-0.20127763$ &    0.01215111 &   1.12448714803832  &1.61545155902268   &1.28160813445288\\
		                s7          &0.6    &   $-0.22973878$ &     0.01214243 &   1.14529304319858 &  1.61314401406252 &1.27974086738477\\
		                s8          &0.4    &  $-0.11133090$ &  0.01177667  &   1.04267905040057& 1.41285715786028  & 1.20385461971527\\
			        s9          &0.4   &    $-0.12329970$& 0.01177641    &   1.04789622892359& 1.41291760105672  &1.20390612161046\\
	                         s10        &0.4    &   $-0.13424207$ &  0.01177617    &    1.05311363730914& 1.41279793409858  & 1.20380415686506\\
	                         s11        &0.4     &   $-0.14439149$  & 0.01177592  &   1.05833035249639&  1.41273744727074 & 1.20375261779282\\
	                         s12        &0.4     &   $-0.16289572$ &   0.01177552  &  1.0687733648426& 1.41264358611283   & 1.20367264142158\\
	                          s13        &0.4     &  $-0.17958219$   &  0.01177309 &  1.07902769262705 &  1.41205766742487 & 1.20317339709575\\
	                           s14        &0.4    &  $-0.19493661$  &   0.01177061  & 1.08927474892939  & 1.41146428986134  &1.2026677972783\\
	                           s15        &0.2     &  $-0.09514373$   &  0.01158044& 1.02015292937479 & 1.26517182375897  &1.16411976115842\\
	                           s16        &0.2    &   $-0.12037734$ &   0.01158021 &  1.03043668219383& 1.26512069418535  &1.16407271541655\\
	                           s17        &0.2     &   $-0.14130345$  & 0.01157971 &  1.04069639252575&  1.2650121280105 &1.16397282066142\\
	                           s18        &0.2     &   $-0.15961623$ &  0.01157844 &  1.05088513370079&   1.26473486730472&1.16371770537942\\
	                           s19        &$-0.2$    & $-0.12027829$   &  $0.01258520$  &1.03041733634713 & 1.16402900625683  &1.26507319082773\\
	                           s20        &$-0.2$    &  $-0.14121083$  & $0.01258398$   & 1.04062105981093 & 1.16380431415062  &1.26482899419842\\
	                           s21        &$-0.2$     & $-0.15948828$  & $0.01257917$   &  1.05052233902211& 1.16291434885511  &1.263861774971\\
	                           s22        &$-0.4$    & $-0.12308769$  &   0.01382007  &1.04783052701748 &   1.20370366381241&1.41267999433539\\
	                           s23        &$-0.4$    &  $-0.14420487$   &  0.01381917 & 1.0582400621434 &   1.20354723275444&1.41249640510764\\
	                           s24        &$-0.4$    &  $-0.16269187$  &  0.01381569 & 1.06844864874712 &  1.20294134980541 &1.41178533414662\\
	                           s25        &$-0.6$    &  $-0.12761312$  &  0.01532347  & 1.08180243941042 & 1.28301992700588  &1.61713969966367\\
	                          s26        &$-0.6$     &  $-0.16815583$  & 0.01531961  &  1.10316051991281&   1.28237359478648&1.61632505176213\\
                                   s27       &$-0.6$    &  $-0.20054310$ &  0.01527546  &  1.12155004323926&   1.27499394723474&1.60702362099378\\
			\end{tabular}
		\end{ruledtabular}
	\end{center}
\end{table*}

\subsection{Initial data}
\label{sec:ID}
The initial data of the Bowen-York form for the two BHs is provided by the {\tt TwoPunctures}~\cite{Ansorg:2004ds}
thorn in the Einstein Toolkit. While the thorn takes as input gauge dependent quantities such 
as initial separation and momenta, we will eventually need gauge-invariant quantities [namely
Arnowitt-Deser-Misner (ADM) energy and angular momentum] to provide meaningful comparisons
with the EOB results, following the same rational of Ref.~\cite{Damour:2014afa}. To have this,
we proceeded as follows.  
The BHs start on the $x$-axis with initial positions designated by $\pm X$ and initial momenta
in the plane $(p_x,p_y)$. The ADM orbital angular momentum is given by
\be
J_{\rm ADM}=2X |p_y|.
\ee

For all simulations we use $X = 50 M$, while $(p_x,p_y,E_{\rm ADM},J_{\rm ADM})$ are listed in
Table~\ref{tab:ID}. For each run, we chose a doublet of $(E_{\rm ADM},J_{\rm ADM})$. In particular,
we keep the same value of $p_y$ for a given spin and then we determine $p_x$ through the iterative
procedure to obtain the target value of $E_{\rm ADM}$. 
That procedure is a root-finding routine wherein we solved for the initial data repeatedly until we  
had the correct input quantities corresponding to those gauge invariant quantities.

Note that, although the data in Table~\ref{tab:ID} are those actually used to start the simulations,
these {\it are not} the quantities of interest to compare the NR with the EOB scattering angle.
To this aim we will need to subtract from the initial ADM quantities the energy and angular momentum
losses due to the junk radiation,  analogously to what done in Ref.~\cite{Damour:2014afa}. 
We shall come back to this below.

\subsection{Mesh refinement and boundary conditions}
\label{sec:grid}
A major challenge of scattering simulations is the need for a very large grid.
Our BHs start $100 M$ apart and end even more widely separated. The large
final separation is partly due to a desire for an accurate angle of deflection.
Additionally, we must run our simulations long enough for radiation from the 
encounter to reach the extraction spheres. 
We ran our simulations on very large computational domains. 
They were all performed in cubes with $(x,y,z)$ coordinates, each 
running from $-5033M$ to $5033M$. In order to accommodate such a large grid
we used 13 layers of box-in-box mesh refinement implemented by the {\tt Carpet} 
thorn~\cite{Schnetter:2003rb,Carpet}.
The outer 6 layers were static, while the inner 7 layers tracked the punctures.
We used radiative boundary conditions although we note that our grid was 
specifically designed to be large enough so reflections off the outer boundary 
did not have time to return and interfere with either the punctures or wave extraction.

\subsection{Calculation of the deflection angle through extrapolation of the BHs trajectories}
\label{sec:traj}
Let us focus first on the description of the procedure adopted to the computation of the
deflection angle, which is our main observable of interest. Since we are limited to a finite-size
grid, the deflection angle is obtained by extrapolating the puncture trajectories to infinity.
For any given run, we perform the following steps:
\begin{enumerate}
\item[(i)] We write the relative motion of the puncture tracks as $(x,y) = (x_1,y_1) - (x_2,y_2)$.
Since the BH spins will always be either aligned or anti-aligned with the orbital angular momentum,
the BHs are never out the $(x,y)$ plane.
\item[(ii)] Convert the relative motion to standard polar coordinates 
$(r,\theta) = (\sqrt{x^2+y^2}, \arctan(y/x))$. 
\item[(iii)] Truncate the $(r,\theta)$ data at both the beginning and the end 
of the simulations. This way we have incoming and outgoing legs representing 
the motion before and after the strong field interaction. We denote these data sets as 
$(r_{\rm in}, \theta_{\rm in})$ and $(r_{\rm out}, \theta_{\rm out})$,
respectively. In practice, we find that a separation of 
$r = 25 M$ is a reasonable cutoff for the strong field. 
\item[(iv)] Perform a least-squares fit on the incoming and outgoing legs 
to find power series representations of $\theta_{\rm in}$ and $\theta_{\rm out}$
 in $1/r$. We repeat this fit for a range of polynomial orders.
We have experimented with several different maximum polynomial orders
and find that cubic and higher orders give consistent results in the final answer.
\item[(v)] Extrapolate each of the incoming and outgoing angles 
to infinite separation by taking the $1/r \to 0$ limits
of the the fits (i.e., take the constant terms in the polynomials),
providing $\theta_{\rm in}^{\infty}$ and $\theta_{\rm out}^{\infty}$.
We choose the extrapolated angles from 4th-order polynomials 
(noting that polynomials of 3rd-6th order give consistent results within our uncertainty ranges).
\item[(vi)] Make (conservative) estimates for the errors in 
$\theta_{\rm in}^{\infty}$ and $\theta_{\rm out}^{\infty}$.
To find the error in $\theta_{\rm in}^{\infty}$ (e.g.)
we first perform our least-squares fit using 1st-4th order polynomials.
We then perform the extrapolation for each polynomial and take the error 
to be the range of the results. 
\item[(vii)] Report the final angle of deflection (in degrees) as 
$\chi = \theta_{\rm out}^{\infty} - \theta_{\rm in}^{\infty} - 180$. 
We take the final error in $\chi$ to be the norm of the 
errors in $\theta_{\rm ins}^{\infty}$ and $\theta_{\rm out}^{\infty}$.
\end{enumerate}
The numerical results for the scattering angles, $\chi^{\rm NR}$, are reported in the
ninth column of Table~\ref{tab:chi_nonspinning} (nonspinning configurations) and in the tenth
column of Table~\ref{tab:chi_spinning} (spinning configurations). The values come with error bars,
that range from a few $\%$'s to well below $1\%$, depending on the dataset.
The total uncertainty estimate is rather conservative and it is done as follows. Each configuration is simulated
at two different resolutions and for each one we compute the scattering angle with the procedure
mentioned above. The finite-difference uncertainty is thus obtained by simply taking 
the difference between the angles obtained with the two resolution. To this value, we additionally
add, in quadrature, the uncertainty related to the extrapolation procedure mentioned above.
Note that for each value of the spin we considered configurations at (approximately)
constant orbital angular momentum and we increased progressively the energy. 
Configuration \#s7 of Table~\ref{tab:chi_spinning} is rather extreme and close to the 
zoom-whirl behavior, since we obtain $\chi^{\rm NR}=371.5(9.9)$~rad. 
We performed multiple simulations with only slightly more energy, each leading to capture.

\begin{table*}[t]
	\caption{\label{tab:chi_nonspinning} Comparison between EOB and NR scattering angle for equal-mass, 
	         nonspinning configurations. The initial angular momentum is kept approximately constant, while the 
	         initial energy is progressively raised. From left to right the columns report:
		the ordering number; the minimum EOB distance $r_{\rm min}^{\rm EOB}$; the initial energy $\hat{E}^{\rm NR}_{\rm in} = E^{\rm NR}_{\rm in}/M$; 
		the initial orbital angular momentum $\hat{J}^{\rm NR}_{\rm in,orb} = J^{\rm NR}_{\rm in,orb}/M^2$;		
		the NR and EOB radiated energies, 
		$(\Delta \hat{E}^{\rm NR},\Delta \hat{E}^{\rm EOB})$; the NR and EOB radiated angular momentum, 
		$(\Delta \hat{J}^{\rm NR},\Delta \hat{J}^{\rm EOB})$; the NR and EOB scattering angles $(\chi^{\rm NR},\chi^{\rm EOB})$ and
		their fractional difference  $\hat{\Delta}\chi^{\rm NREOB}\equiv |\chi^{\rm NR}-\chi^{\rm EOB}|/\chi^{\rm NR}$. }
	\begin{center}
		\begin{ruledtabular}
			\begin{tabular}{c c c c  c c c c c c c } 
				$\#$  & $r_{\rm min}^{\rm EOB}$ & $\hat{E}^{\rm NR}_{\rm in}$ & $\hat{J}^{\rm NR}_{\rm in,orb}$ & $\Delta \hat{E}^{\rm NR}$ & $\Delta \hat{E}^{\rm EOB}$ & $\Delta \hat{J}^{\rm NR}$ & $\Delta \hat{J}^{\rm EOB}$ & $\chi^{\rm NR}$ [deg] & $\chi^{\rm EOB}$[deg] & $\hat{\Delta}\chi^{\rm NREOB}[\%]$ \\
				\hline
				1& 6.70 & 1.0045678(42) & 1.1520071(73) & 0.001625(16) & 0.0014 & 0.034511(71) & 0.0278 & 201.9(4.8) & 200.5173 & 0.69 \\		
				2& 5.02 & 1.0147923(76)  & 1.151918(16)    & 0.004925(30) & 0.0042 & 0.069504(39) & 0.0549 & 195.9(1.3) & 194.5465 & 0.71 \\
				3& 4.46 & 1.0198847(82) & 1.151895(11)   & 0.00793(34) & 0.0068 & 0.09456(70) & 0.0757 & 207.03(99) & 207.0698 & 0.02 \\
				4& 3.97 & 1.024959(12)   & 1.151845(12)    & 0.01194(27) & 0.0109 & 0.1252(10) & 0.1055 & 225.54(87) & 229.0237  & 1.54 \\
				5& 3.13 & 1.035031(27)   & 1.1515366(78) & 0.0281(11)     & 0.0306 & 0.2220(64) & 0.2283 & 307.13(88) & 345.9284  & 12.63
			\end{tabular}
		\end{ruledtabular}
	\end{center}
\end{table*}

\begin{table*}
	\caption{
		\label{tab:chi_spinning}
Comparison between EOB and NR scattering angle for equal-mass, equal-spin BBHs. The column entries are the same of Table~\ref{tab:chi_nonspinning}, with the addition 
of the spin parameter $a/M$. For each dimensionless spin value of $S_{1,2}/m_{1,2}^2$, the initial angular momentum is almost constant while the energy is changed.
When {\tt TEOBResumS} predicts a plunge instead of a scattering, the EOB quantities cannot be meaningfully compared and are hence substituted by dots. 
This happens for the most energetic systems.
	}
	\begin{ruledtabular}
		\begin{tabular}{c c c c c c c c c c c c} 
		$\#$ &$S_{i}/m_{i}^2$ & $r_{\rm min}^{\rm EOB}$ & $\hat{E}^{\rm NR}_{\rm in}$ & $\hat{J}^{\rm NR}_{\rm in,orb}$ & $\Delta \hat{E}^{\rm NR}$ & $\Delta \hat{E}^{\rm EOB}$ & $\Delta \hat{J}^{\rm NR}$ & $\Delta \hat{J}^{\rm EOB}$ & $\chi^{\rm NR}$ [deg] & $\chi^{\rm EOB}$[deg] & $\hat{\Delta}\chi^{\rm NREOB}[\%]$ \\
			\hline 
			s1  & 0.6  & 3.25 & 1.076631(25)  & 1.283562(22)   & 0.004530(90) & 0.009  & 0.06944(42)   & 0.0835 & 153.1(1.3) & 157.2555 & 2.70 \\
			s2  & 0.6  & 3.09 & 1.081965(13)  & 1.283473(27)   & 0.00636(16)  & 0.0109 & 0.08527(37)   & 0.0954 & 155.6(1.1) & 165.6316 & 6.43 \\
			s3  & 0.6  & 2.94 & 1.087309(15)  & 1.2834073(65)  & 0.008449(88) & 0.0134 & 0.10285(95)   & 0.1103 & 159.9(1.1) & 175.4255 & 9.73 \\
			s4  & 0.6  & 2.79 & 1.092679(90)  & 1.283300(23)   & 0.01098(18)  & 0.0169 & 0.1223(16)    & 0.1301 & 165.6(1.1) & 187.5039 & 13.22 \\
			s5  & 0.6  & 2.51 & 1.10326(12)   & 1.283004(13)   & 0.0178(11)   & 0.0305 & 0.1666(24)    & 0.1992 & 181.36(91) & 227.0854 & 25.21 \\
			s6  & 0.6  & $\cdots$ & 1.124160(89)  & 1.281608(41)   & 0.0410(24)   & $\cdots$ & 0.282(12)     & $\cdots$ & 234.0(1.4) & $\cdots$ & $\cdots$ \\
			s7  & 0.6  & $\cdots$ & 1.14467(26)   & 1.279741(39)   & 0.084(11)    & $\cdots$ & 0.442(35)     & $\cdots$ & 371.5(9.9) & $\cdots$ & $\cdots$ \\
			\hline
			s8  & 0.4  & 3.88 & 1.042630(24)  & 1.2038568(74)  & 0.00501(12)  & 0.0076 & 0.07063(42)   & 0.0745 & 163.4(1.2) & 164.5544 & 0.69 \\
			s9  & 0.4  & 3.62 & 1.047832(16)  & 1.2038078(17)  & 0.00693(11)  & 0.0097 & 0.08809(66)   & 0.0888 & 168.0(1.0) & 174.4029 & 3.78 \\
			s10 & 0.4  & 3.38 & 1.053050(62)  & 1.203739(10)   & 0.00957(14)  & 0.0125 & 0.1080(11)    & 0.106  & 174.95(91) & 186.3945 & 6.54 \\
			s11 & 0.4  & 3.16 & 1.058213(92)  & 1.2036948(54)  & 0.01291(13) & 0.0162 & 0.1303(33)    & 0.1272 & 183.92(79) & 200.9462 & 9.26 \\
			s12 & 0.4  & 2.74 & 1.068602(39)  & 1.20361881(22) & 0.02088(32)  & 0.0287 & 0.1834(46)    & 0.1942 & 209.12(92) & 246.5841 & 17.92 \\
			s13 & 0.4  & $\cdots$ & 1.078749(45)  & 1.2031336(88)  & 0.0359(25)   & $\cdots$ & 0.2495(83)    & $\cdots$ & 248.76(85) & $\cdots$ & $\cdots$ \\
			s14 & 0.4  & $\cdots$ & 1.088934(80)  & 1.2026022(50)  & 0.0571(60)   & $\cdots$ & 0.345(26)     & $\cdots$ & 322.6(2.5) & $\cdots$ & $\cdots$ \\
			\hline
			s15 & 0.2  & 4.86 & 1.0201378(96) & 1.1640845(68)  & 0.003958(93) & 0.0041 & 0.059581(34)  & 0.0518 & 175.7(1.4) & 171.7437 & 2.25 \\
			s16 & 0.2  & 4.02 & 1.030376(18)  & 1.164026(11)   & 0.00877(15)  & 0.0086 & 0.09980(80)   & 0.085  & 189.30(91) & 191.8992 & 1.37 \\
			s17 & 0.2  & 3.35 & 1.040580(25)  & 1.1639250(95)  & 0.01649(32)  & 0.0176 & 0.1558(33)    & 0.1421 & 218.89(83) & 232.1026 & 6.04 \\
			s18 & 0.2  & 2.77 & 1.050656(99)  & 1.163691(10)   & 0.0326(20)   & 0.04   & 0.2379(99)    & 0.2649 & 276.65(65) & 334.3757 & 20.87 \\
			\hline
			s19 & $-0.2$ & 4.4  & 1.0303638(21) & 1.2650314(44)  & 0.006533(51) & 0.0085 & 0.08797(20)   & 0.0912 & 177.67(70) & 188.4583 & 6.07 \\
			s20 & $-0.2$ & 3.55 & 1.040516(17)  & 1.2647776(59)  & 0.014385(25) & 0.0215 & 0.147468(20)  & 0.1813 & 212.4(3.6) & 252.3961 & 18.83 \\
			s21 & $-0.2$ & $\cdots$ & 1.050347(27)  & 1.2638274(88)  & 0.031533(42) & $\cdots$ & 0.248076(70)  & $\cdots$ & 292.9(2.0) & $\cdots$ & $\cdots$ \\
			\hline
			s22 & $-0.4$ & 4.13 & 1.047768(19)  & 1.4126441(60)  & 0.004003(88) & 0.014  & 0.06741(25)   & 0.1363 & 148.01(77) & 183.1852 & 23.76 \\
			s23 & $-0.4$ & 3.35 & 1.058124(25)  & 1.412432(10)   & 0.007071(36) & 0.0346 & 0.085915(12) & 0.2739 & 168.9(2.7) & 268.3712 & 58.91 \\
			s24 & $-0.4$ & $\cdots$ & 1.068236(32)  & 1.411707(16)   & 0.017072(47) & $\cdots$ & 0.165944(20)  & $\cdots$ & 206.6(3.4) & $\cdots$ & $\cdots$ \\
			\hline
			s25 & $-0.6$ & 3.25 & 1.081701(28)  & 1.6170091(50)  & 0.002527(38) & 0.0494 & 0.0538724(65) & 0.3844 & 125.5(3.3) & 266.2012 & 112.15 \\
			s26 & $-0.6$ & $\cdots$ & 1.102951(38)  & 1.616151(26)   & 0.010602(56) & $\cdots$ & 0.1400586(54)  & $\cdots$ & 160.5(2.6) & $\cdots$ & $\cdots$ \\
			s27 & $-0.6$ & $\cdots$ & 1.121146(67)  & 1.606899(38)   & 0.043455(89) & $\cdots$ & 0.346465(94)  & $\cdots$ & 271.5(3.0) & $\cdots$ & $\cdots$
		\end{tabular}
	\end{ruledtabular}
\end{table*}

\subsection{Gravitational waveforms and fluxes}
\label{sec:wave}
The other quantities of interest are the gravitational waveform and the fluxes. The control of the energy fluxes
is needed to meaningfully match the NR results with the EOB ones, i.e. to start the EOB evolution consistently
with the NR simulations. We computed the Weyl scalar $\psi_4$ at 10 radii, ranging from 260$M$  to 500$M$,
evenly spaced in $1/r$. The Weyl scalar was decomposed into spin-weighted
spherical harmonics including all harmonics up to $\ell = 8$. In order to find $\psi_4$ at infinity, we 
extrapolated each of these multipoles using the {\tt SimulationTools}~\cite{SimulationTools} {\tt Mathematica} 
package. From $\psi_4$ we computed the strain, and out of it the total energy and angular momentum radiated
during the evolution. The total losses $(\Delta E^{\rm NR},\Delta J^{\rm NR})$ reported in Tables~\ref{tab:chi_nonspinning} 
and~\ref{tab:chi_spinning} are obtained including all multipoles up to $\ell=8$. Note that the resulting finite-radius 
values of $(\Delta E^{\rm NR},\Delta J^{\rm NR})$ are then extrapolated to null infinity. 
Analogously to the case of the scattering angle, we give a conservative estimate of the uncertainty due to finite-difference 
error by computing the fluxes for the two different resolutions available and then taking the difference.

\begin{figure}[t]
\includegraphics[width=0.50\textwidth]{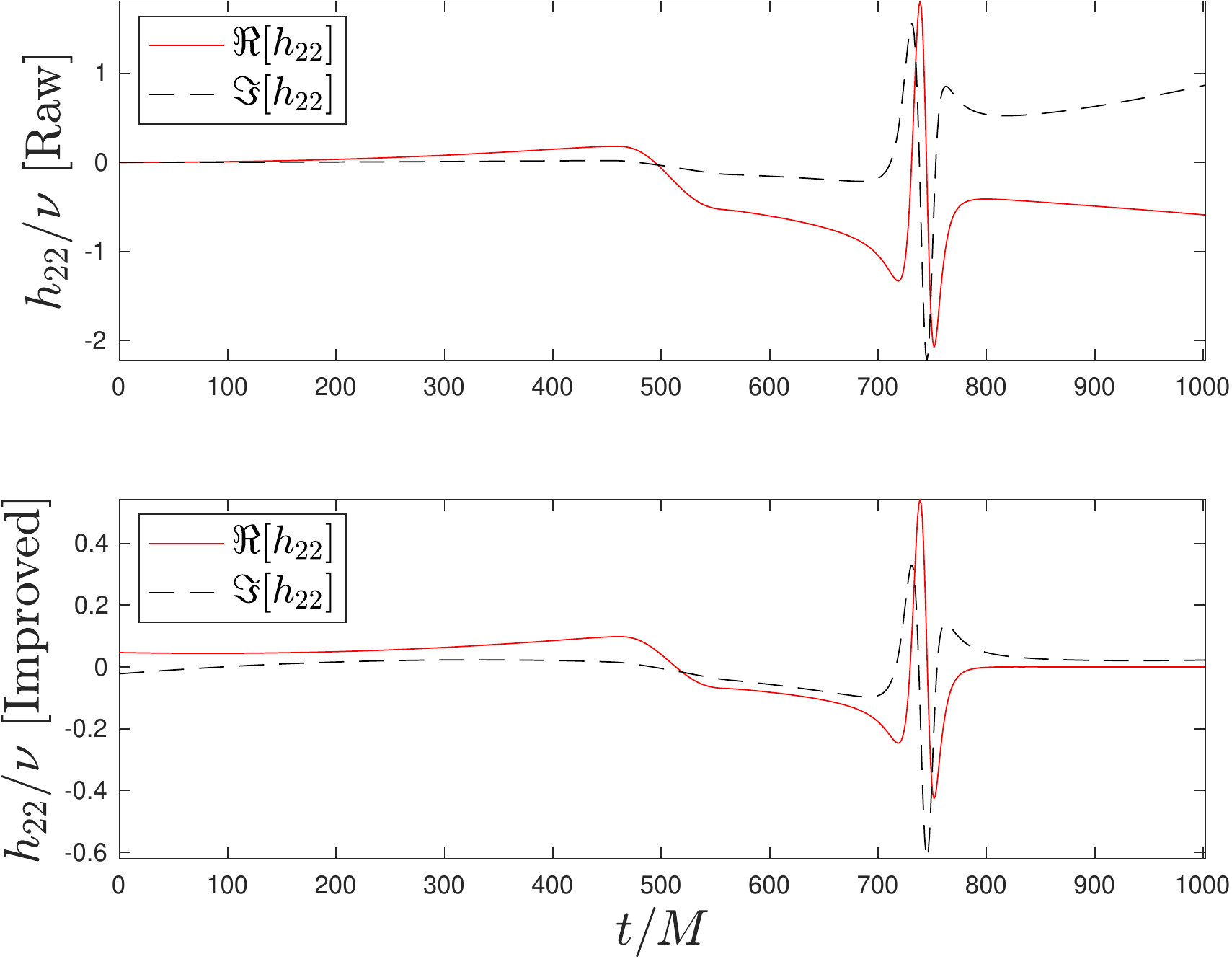}
\caption{
\label{fig:h_s04}
Obtaining the strain $h_{22}$ from $\psi^4_{22}$  using double integration in the time-domain, Eq.~\eqref{eq:psi_to_h},
demonstrated on the waveform from configuration $\#4$ of Table~\ref{tab:chi_nonspinning}. 
Top panel: direct double integration in the time-domain, that highlights the presence of an unphysical global drift on 
both the real and imaginary parts. Bottom panel: improved waveform obtained removing a global straight line fitted to
the data after each time-domain integration.}
\end{figure}

In this respect, it is worth mentioning that there is an especially delicate (and in fact dominant) source of uncertainty 
in the calculation of the angular momentum flux that we {\it do not} consider in our error bars. This comes from
the recovery of the strain from the Weyl scalar $\psi_4$, a procedure  that needs two time-integrations and thus 
the determination of two, arbitrary, integration constants. This turns out to be a complicated procedure 
for our simulations and we have investigated it briefly only for the $\ell=m=2$ mode. 
We use the following convention for the metric strain
\be
h_+ - i h_\times = {\cal D}_L^{-1}\sum_{\ell=2}^{\rm \ell_{\rm max}}\sum_{\ell=-m}^m h_\lm {}_{-2} Y_\lm \ ,
\ee
where ${\cal D}_L$ is the luminosity distance and ${}_{-2} Y_\lm $ are the $s=-2$ spin-weighted
spherical harmonics. The usual relation between the strain modes and the $\psi^4$ modes
is given by $\ddot{h}_\lm = {\cal D}_L \psi^4_\lm$. The $\ell=m=2$ strain mode is thus
obtained as
\begin{align}
\label{eq:psi_to_h}
h_{22} = \int_{0}^t dt' \int_{0}^{t'} dt'' {\cal D}_L\psi^4_{22}.
\end{align}
It is well known that the application of this formula on numerical waveforms extracted at finite radius may
give unphysical features, with global drifts that might be nonlinear and difficult to subtract~\cite{Berti:2007fi}.
For quasi-circular binaries, performing the integration in the frequency domain typically gives a robust
result~\cite{Reisswig:2010di}. The scattering, or hyperbolic encounter, setup doesn't seem to have been 
explored so far in the literature. Notably Ref.~\cite{Gold:2012tk} only explicitly shows $\psi^4_\lm$ 
waveforms and did not attempt to compute the metric strain. Figure~\ref{fig:h_s04} refers to configuration
$\#4$ of Table~\ref{tab:chi_nonspinning}. The top panel of the figure shows the result of the direct integration
of $\psi^4_{22}$: both the real and imaginary part of the waveform exhibit a quasi-linear, global, unphysical
drift. Subtracting it by just removing a linear regression doesn't seem to work. By contrast, in the bottom panel of the figure
we show an improved waveform that is obtained by subtracting a global linear fit after {\it each} integration.
The interval where the fit is performed is $[900,1000]$ for the real part and $[2,1000]$ for the imaginary part.
Although the waveform looks qualitatively correct, the fact that the global quadratic behavior of the imaginary
part is still indicating that some systematics is present. In general, this is not surprising since the main aim
of our simulations was the computation of the angle from the puncture tracks, and not specifically the computation
of the waveforms. It is likely that different grid setup, resolution, extraction radii could be useful to reduce 
the waveform uncertainties. Despite this, there are a few considerations that is still worth making using the 
current data. First, one notes that the waveform as a sort of bump around $t=400M$, that is prominent in the 
real part while it is absent in the imaginary part. That feature is related to the initial burst of junk radiation that is present
in $\psi^4_{22}$ at approximately the same time. 
Second, as already mentioned above, differences in the strain will reflect on the computation 
of  the losses $(\Delta J^{\rm NR},\Delta E^{\rm NR})$  listed in Tables~\ref{tab:chi_nonspinning}-\ref{tab:chi_spinning}. 
Focusing only on the (dominant) $\ell=m=2$ waveform mode we computed the losses using either the raw waveform, 
getting $(\Delta J^{\rm NR}_{22},\Delta E^{\rm NR}_{22})=(0.011571,0.12165)$, or the {\it improved} waveform, 
obtaining $(\Delta J^{\rm NR}_{22},\Delta E^{\rm NR}_{22})=(0.011566,0.11922)$. If the fractional difference for
$\Delta E^{\rm NR}_{22}$ is negligible, for $(\Delta J^{\rm NR}_{22}$ we get a $2\%$, that might be even slightly 
larger because of the behavior of the precursor. In conclusion, our analysis shows that the 
calculation of $\Delta J^{\rm NR}$ could be overestimated by a few percents due to the double integration procedure.
This specific analysis, as well as discussion of waveforms, is not within the scopes of this analysis and it is thus 
postponed to future work.

\section{EOB/NR comparisons}
\label{sec:eobnr}

Reference~\cite{Damour:2014afa} pioneered the EOB/NR comparison of the scattering angle using
several flavors of a (now outdated) EOB model. It is worth stressing that at the time a model 
incorporating radiation reaction along generic orbits was not available, so that it was possible instead
to use the NR losses using a formula proposed in Ref.~\cite{Bini:2012ji} that neglects quadratic effects in the
radiation reaction and only relies on the knowledge of the EOB Hamiltonian.
More recently, the progressive development of faithful EOB models valid for generic, non-quasi-circular, 
configurations~\cite{Nagar:2020xsk,Nagar:2021gss}, and in particular the use of a consistent radiation
reaction force, has revived the interest for scattering configurations so to test the model in an extreme 
regime. In particular, Ref.~\cite{Nagar:2021gss} showed that an EOB model informed by quasi-circular
NR simulations yields values of the scattering angle that are more consistent, also in strong field, with 
the NR computations of Ref.~\cite{Damour:2014afa}.
This EOB/NR agreement was improved further in Ref.~\cite{Nagar:2021xnh}, that proposed an improved
version of the \TEOBResumS{} model for spin-aligned binaries on generic orbits crucially incorporating 
high-order (nonspinning)  PN information~\cite{Bini:2019nra,Bini:2020wpo,Bini:2020nsb,Bini:2020hmy} 
within the  EOB potentials. In particular, Table~V of Ref.~\cite{Nagar:2021xnh} reports what is currently
the best agreement between EOB and NR scattering angles, with a fractional difference $\sim 3.8$~\%
for the most relativistic configuration computed numerically, reaching an EOB impact parameter $\sim 3.5$.
\begin{figure}[t]
	\center
	\includegraphics[width=0.47\textwidth]{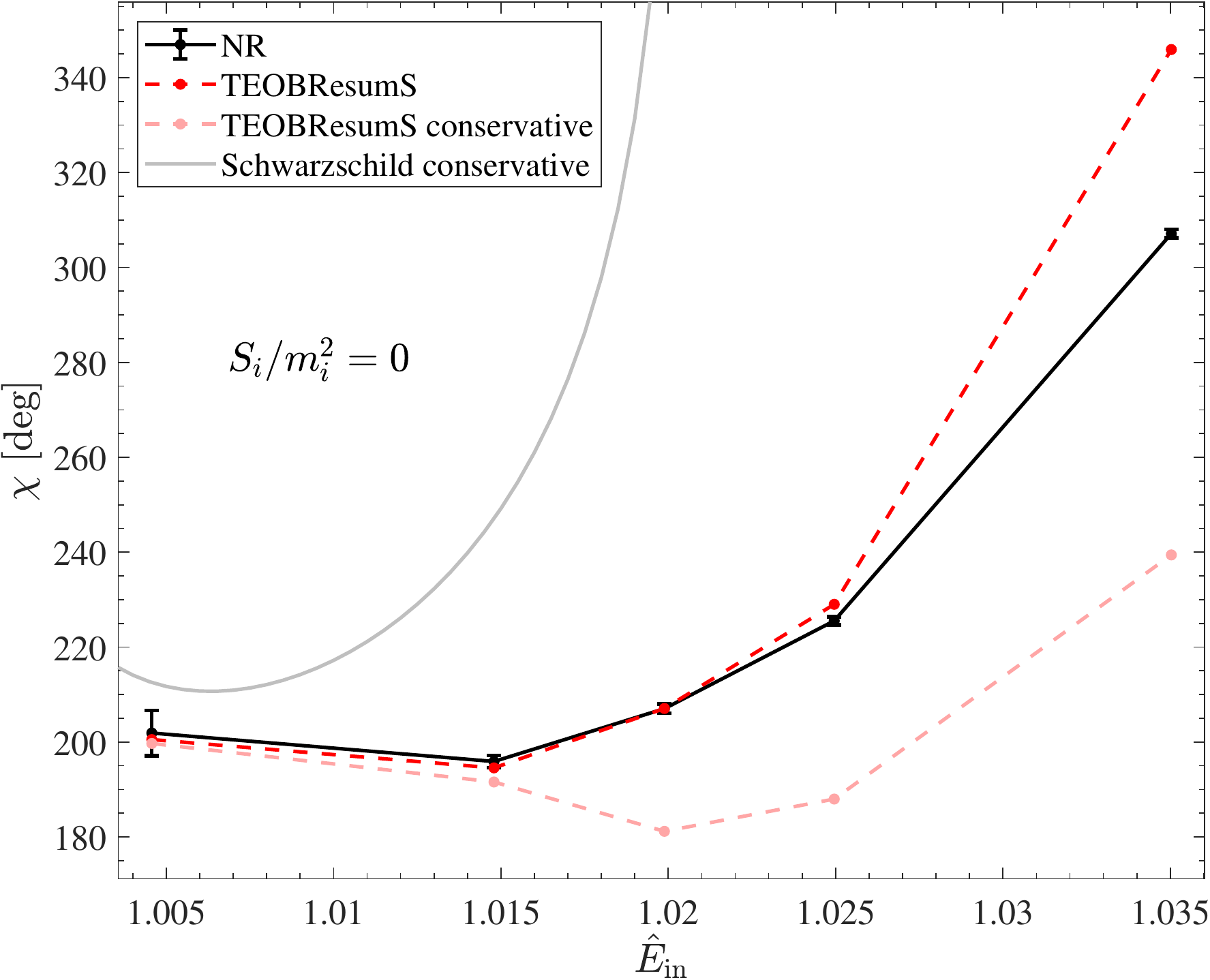}
	\caption{\label{fig:chi_s0}		
		Scattering angle versus initial energy for the nonspinning systems of Table~\ref{tab:chi_nonspinning} 
		for NR simulations (solid black curve) and \TEOBResumS{} (dashed red line).
		We also plot the \TEOBResumS{} scattering angle in the case of conservative motion (dashed pink curve) to show the impact of radiation reaction.
		The same is done for the case of a test particle moving around a Schwarzschild BH (grey line).
		Note that the the Schwarzschild line is computed using an initial angular momentum $p_\varphi^{\rm Schw}$ corresponding to the $\hat{J}^{\rm NR}_{\rm in, orb}$ to the first simulation of Table~\ref{tab:chi_nonspinning} ($\#1$).
		We can see that \TEOBResumS{} is consistent with NR simulations for smaller energies but starts to deviate at higher ones.		
	}
\end{figure}
For our current comparison we thus use the model of Ref.~\cite{Nagar:2021xnh}. In order to ease the discussion, 
we do not report here any technical detail about the EOB construction, but directly refer the reader to Ref.~\cite{Nagar:2021xnh}.
As mentioned above, in order to assure a consistent match between EOB and NR simulations, the 
EOB dynamics is started with the NR energy 
and angular momentum $(\hat{E}^{\rm NR}_{in},\hat{J}^{\rm NR}_{\rm in, orb})$ calculated {\it after} the junk radiation, following precisely the scheme 
of Ref.~\cite{Damour:2014afa}. We fix the EOB initial separation to $r_0=10000$; then, we 
identify $(\hat{E}^{\rm NR}_{in},\hat{J}^{\rm NR}_{\rm in, orb})$  with the initial EOB energy and 
angular momentum and out of this we determine the initial radial momentum $p_{r_*}$.
From the evolution we compute the orbital phase $\varphi$ as function of time and we finally obtain $\chi^{\rm EOB}=\varphi-\pi$.

\begin{figure}[t]
	\center
	\includegraphics[width=0.47\textwidth]{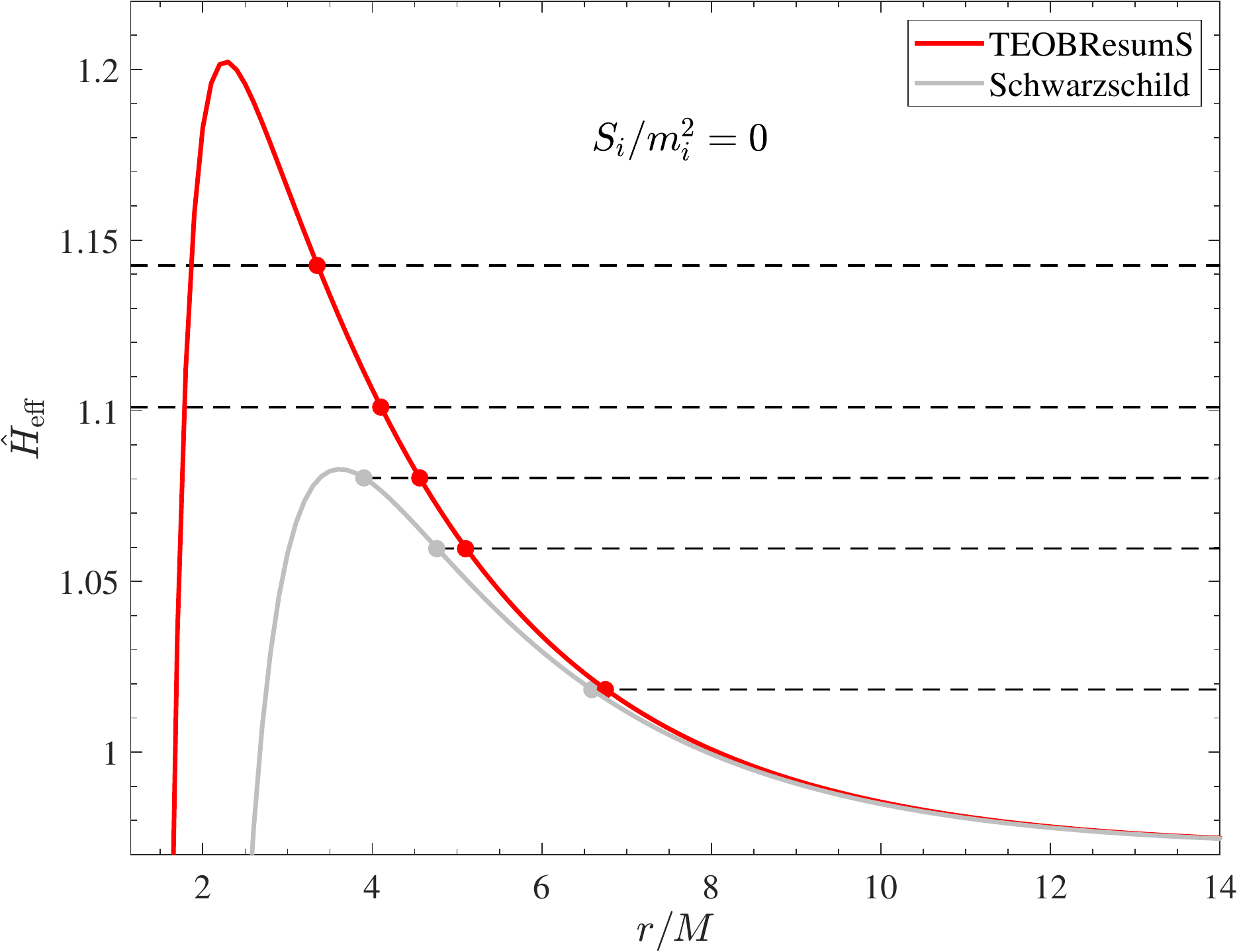}
	\caption{\label{fig:H0}  
		Comparison between the circular \TEOBResumS{} effective Hamiltonian for equal-mass, nonspinning systems (red line) with its Schwarzschild equivalent (grey line).
		The angular momentum is set to the value of $\hat{J}^{\rm NR}_{\rm in,orb}$ of the first NR simulation of Table~\ref{tab:chi_nonspinning}, so neglecting the (small) 
		differences in initial angular momenta between the various configurations.
		The dotted horizontal lines correspond to the initial effective energies $\hat{E}_{\rm eff, in}^{\rm NR}$, computed from the $\hat{E}_{\rm in}^{\rm NR}$ of 
		Table~\ref{tab:chi_nonspinning} using Eq.~\eqref{eq:Eeff}. We can see how these systems scatter within \TEOBResumS{}, since the energies are all 
		lower than the potential barrier, while this is not always the case for the Schwarzschild potential.
	}
\end{figure}

We focus first on the nonspinning configurations, that are reported in Table~\ref{tab:chi_nonspinning}. 
For completeness, the data are also shown in Fig.~\ref{fig:chi_s0}. The model has an outstanding agreement with 
the simulated systems for smaller energies, which progressively worsens with the increase of the initial energy.
This is to be expected, as high energies imply smaller impact parameters and hence strong-field interactions.
We also show in Fig.~\ref{fig:chi_s0} the scattering angle obtained using {\it only} the conservative sector of the EOB dynamics,
setting to zero the radiation reaction force. The difference is noticeable, especially as the energy
increases, which highlights the importance of the analytical radiation reaction in yielding consistent
EOB/NR scattering angles.
To orient the reader, we also report in Fig.~\ref{fig:chi_s0} the scattering angle of a particle moving
along the geodesics of the Schwarzschild metric.
In this latter case, we identify $p_\varphi^{\rm Schw}=\hat{J}^{\rm NR}_{\rm in,orb}/\nu$
and $\hat{E}^{\rm Schw} = \hat{E}_{\rm eff, in}^{\rm NR}$, where the dimensionless effective energy is computed through the EOB relation 
\begin{equation}
\label{eq:Eeff}
	\hat{E}_{\rm eff} \equiv \frac{E^2 - m_1^2 - m_2^2}{2 m_1 m_2}.
\end{equation}
Finally, we can visualize the energy regime we are exploring by looking at the initial circular Hamiltonians. 
We remind the reader that the real EOB Hamiltonian is obtained from the effective one through
Eq.~\eqref{eq:Eeff} and reads
\begin{equation}
	H_{\rm EOB} = M \sqrt{1 + 2\nu\left(\hat{H}_{\rm eff} - 1\right)}.
\end{equation}
while the effective Hamiltonian (per unit of reduced mass) reads
\be
\hat{H}_{\rm eff}\equiv \dfrac{H_{\rm eff}}{\mu}=\sqrt{p_{r_*}^2 + A\left(1+p_\varphi^2 u^2)+ Q\right)}\ ,
\ee
where $(A,Q)$ are the EOB potentials at 5PN (formal) accuracy as discussed in Ref.~\cite{Nagar:2021xnh}.
In the Schwarzschild limit, $Q=0$ and $A=1-2 u$. In order to have a meaningful comparison, in Fig.~\ref{fig:H0} 
we show the Schwarzschild and \TEOBResumS{}  effective  Hamiltonians together with the effective 
initial energies $\hat{E}_{\rm eff, in}^{\rm NR}$.

\begin{figure}[t]
	\center
	\includegraphics[width=0.47\textwidth]{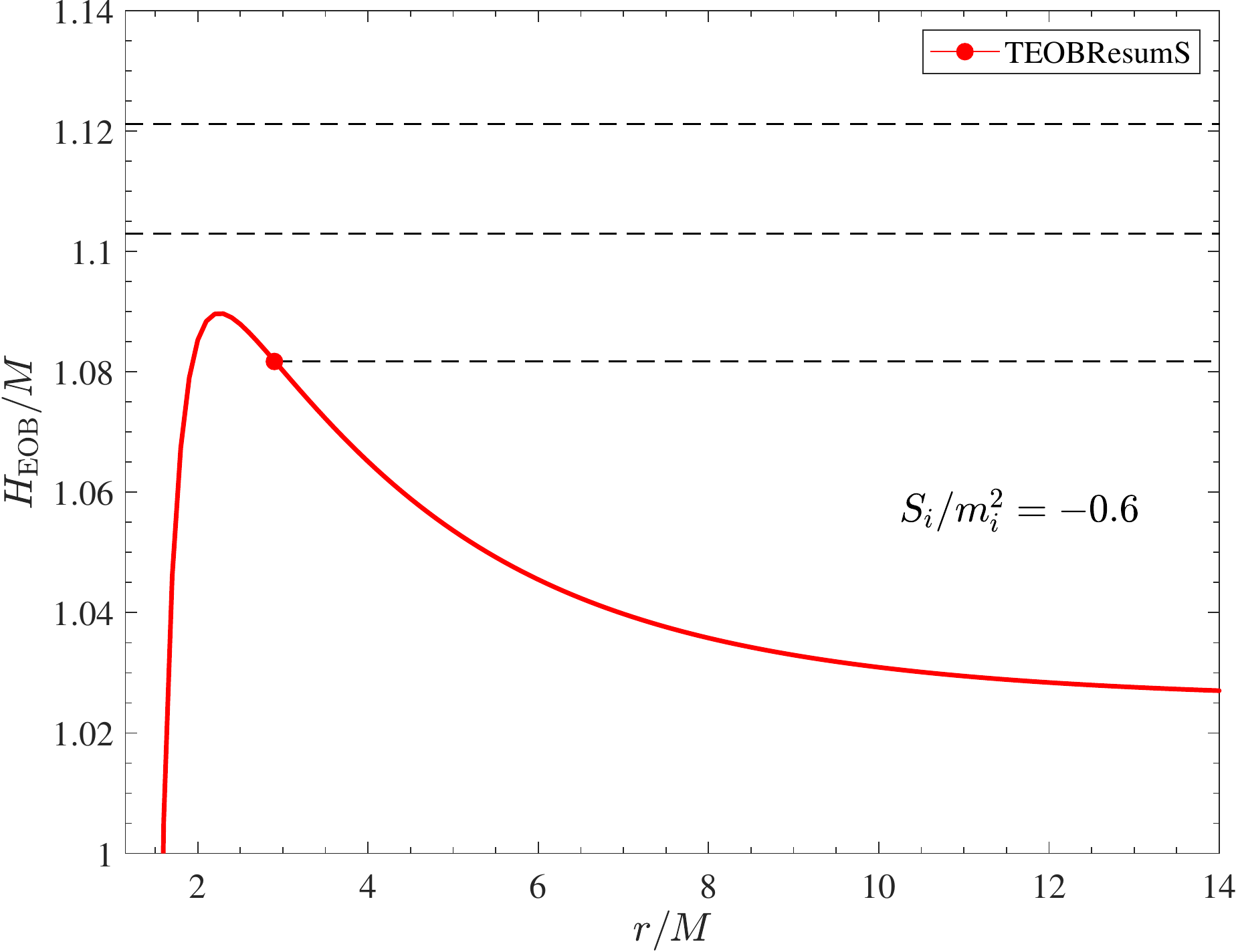}
	\caption{\label{fig:Hm06} 
		\TEOBResumS{} circular Hamiltonian for equal-mass systems with large, anti-aligned spins ($S_i/m_i^2 = -0.6$).
		The angular momentum is set to the value of $\hat{J}^{\rm NR}_{\rm in,orb}$ of the first NR simulation with the same spin in Table~\ref{tab:chi_spinning} ($\#s25$).
		The dotted horizontal lines correspond to the initial energies $\hat{E}_{\rm in}^{\rm NR}$ of the last 3 systems in Table~\ref{tab:chi_spinning}.
		Here we note how two of the systems are plunging instead of scattering, because the potential barrier of \TEOBResumS{} is too low.
	}
\end{figure}


Spinning configurations are listed in Table~\ref{tab:chi_spinning}. One sees that the EOB/NR consistency is of the order
of a few percents for the lowest energy configurations and for spins aligned with the orbital angular momentum.  
By contrast, when the spins are anti-aligned with the orbital angular momentum, the EOB/NR differences are rather
large even for the smallest values of the energy. For large energies, while the NR simulations give a scattering, the
corresponding EOB dynamics gives a plunge. The same also occurs for the spin-aligned case for the largest values
of the energy. Phrasing this phenomenon in a different way, it seems that the EOB gravitational interaction is {\it more attractive}
than what it is supposed to be. Not that although this is very evident for high energies, the global qualitative behavior is the
same for all configurations, with the EOB angle always overestimating the NR prediction. This suggests that
the strong-field behavior of the eccentric version of {\tt TEOBResumS}, and possibly the spin sector of the
model, should be improved in some way to better match these configurations.
In practical terms, it is instructive to look at the EOB circular Hamiltonian for $S_i/m_i^2=-0.6$, Fig.~\ref{fig:Hm06}. 
The three horizontal lines are the energies for the three configurations with the same spin values in Table~\ref{tab:chi_spinning}. 
It is evident from this plot that to have a scattering for configurations $\#s26$ and $\#s27$ the EOB Hamiltonian should 
peak at a much higher value, yielding a larger potential barrier. The understanding of the modifications needed 
to the spin-sector of the current \TEOBResumS{} eccentric model is postponed to future work.
Let us only mention that it will be interesting to explore the performance of the new description of the spin-orbit 
sector proposed in the concluding section of Ref.~\cite{Rettegno:2019tzh}, that uses a different gauge choice 
to incorporate, in simple form, the Hamiltonian of a spinning particle derived in Ref.~\cite{Barausse:2009aa}.

\section{Conclusions}
\label{conclusions}
We have performed new numerical computations of the scattering angle $\chi$ of equal-mass BBH encounters.
For the first time we have considered spinning BHs, with spins aligned or anti-aligned with the orbital angular momentum. This complements the 10, nonspinning, configurations simulated in 
Ref.~\cite{Damour:2014afa} with other 32  configurations, increasing the NR knowledge of the parameter space of scattering BBHs.
Differently from the results of Ref.~\cite{Damour:2014afa}, that were performed at fixed
incoming energy, here we keep the initial angular momentum fixed while the incoming energy is varied.
We find that the magnitude of the scattering angle increases with the incoming energy, analogously to the behavior 
of a test-particle scattering on a Kerr BH.  The numerical results are compared with the predictions of the 
most developed version of the eccentric  \TEOBResumS{} model, introduced in Ref.~\cite{Nagar:2021xnh}.
For the nonspinning case, the new simulations further confirm the quality of the model also for scattering configurations, 
as already pointed out in Ref.~\cite{Nagar:2021xnh}. The EOB/NR agreement is $\lesssim 1\%$ in all cases except
for configuration $\#5$ of Table~\ref{tab:chi_nonspinning} where the EOB angle is larger  than NR
prediction by $\sim 13\%$. This indicates that the EOB dynamics should be improved further, notably by reducing the attractive
character of the EOB potential $A$ [e.g. pushing the effective horizon at smaller values of $r$ by suitably tuning the effective 
5PN parameter $a_5^c(\nu)$] so to consistently reduce the magnitude of the scattering angle. This result is a priori not 
surprising, since the potential $A$ was NR-informed only using quasi-circular NR simulations, i.e. in a rather different 
physical setup. Nonetheless, the $\simeq 13\%$ difference found in the most relativistic configuration actually
indicates that the NR-tuning of $A$, as discussed in Ref.~\cite{Nagar:2021xnh}, is robust and only partially 
dependent on the specific physical setup chosen to do it (in this case, quasi-circular configurations).
By contrast, as far as spins are concerned the EOB/NR differences grow considerably, typically yielding scattering
angles that are {\it larger} than the NR predictions. This indicates that the global attractive character of the EOB potentials
(as obtained from the combined action of orbital and spin-orbital interactions) is too strong and should be weakened.
How to do so in practice will hopefully be explored in future work.

\begin{acknowledgments}
We thank S.~Albanesi for help to correctly obtain the strain waveform of Fig.~\ref{fig:h_s04}.
S.H. and A.N. acknowledge useful discussions with H.~Pfeiffer and H.~R\"uter during the workshop {\it Gravitational scattering,
inspiral and radiation}, Galileo Galilei Institute, April 19 2021--May 21 2021, where this work was presented for the first time. 
We also thank Ian Hinder and Barry Wardell for the SimulationTools analysis package that we used to analyze our NR simulations. 
Simulations were performed on supercomputers at the Albert Einstein Institute.
\end{acknowledgments}

\bibliography{refs20220422.bib,local.bib}

\end{document}